\documentclass[10pt,letterpaper]{article}

\usepackage{kashsty}
\usepackage{kashthm}
\usepackage[round,authoryear]{natbib}

\usepackage[margin=1.37in]{geometry}

\def\PICDIR{.}

\title{
	Local Statistics for Spatial Panel Models 
	with Application to the US Electorate
}
\author{
	Jianfeng Wang,
	Adam B Kashlak\\
	Mathematical \& Statistical Sciences\\
	University of Alberta\\
	Edmonton, Canada,  T6G 2G1}

\begin{document}
	
	\maketitle
	
	\begin{abstract}
		The spatial panel regression model has shown great 
		success in 
		modelling econometric and other types of data that
		are observed both spatially and temporally with 
		associated predictor variables.  
		However, model checking via testing for spatial correlations
		in spatial-temporal residuals is still lacking. 
		We propose a general methodology for fast permutation
		testing of local and global indicators of spatial 
		association. This methodology extends past statistics
		for univariate spatial data 
		that can be written as a 
		gamma index for matrix similarity to the 
		multivariate and panel data settings. 
		This includes Moran's $I$ and
		Geary's $C$ among others.
		Spatial panel models are fit and our methodology is 
		tested on county-wise electoral results for the 
		five US presidential elections from 2000 to 2016
		inclusive.
		County-wise exongenous predictor variables included in this 
		analysis are voter population density, median income,
		and percentage of the population that is non-hispanic
		white.
	\end{abstract}

\maketitle

\section{Introduction}

Few datasets are more intriguing than those surrounding
national elections.  Trying to determine which factors 
influence the whims of the electorate is of importance
for both historical study and future predictions.
Political and demographic data naturally exists 
in a spatial temporal setting.  One tool 
for modelling such data is the spatial panel 
regression model, which combines panel data
analysis---i.e. where measurements are taken on 
the same units over an extended time period---with a 
spatial dependency network.  Details are references are
in Section~\ref{sec:spModel}.

Beyond mere modelling of political performance 
using spatial panel models, we are interested in 
the model's performance in fitting to the data.
To this end, we extend a variety of 
local indicators of spatial association (LISA)
to apply to the residuals produced by 
spatial panel models.  Furthermore, these test 
statistics are evaluated via a quick to compute
analytic variant of the classic permutation test
with details and references in 
Section~\ref{sec:permTest}.
This allows for fast exact testing of regression
residuals for local spatial association across 
a large graphical network.
We will test our methods on a data set of 5 years $\times$ 
3104 counties worth of electoral and demographic data.

The US elections and demographics data is introduced in 
Section~\ref{sec:usData}.  Some regressions and plots are
displayed to give a picture of the data under 
consideration.
A variety of multivariate LISA statistics are 
introduced in Section~\ref{sec:lisaStats}
and non-asymptotic exact significance testing of
such statistics is discussed in Section~\ref{sec:sigTest}.
A simulation study on a $50\times60$ rectangular grid
is performed in Section~\ref{sec:simStudy} to compare
four LISA statistics on multivariate Gaussian data.
Fitted spatial panel models on the US elections data
are detailed in Section~\ref{sec:realSPLM}.
LISA tests for the US data are performed in 
Section~\ref{sec:realLISA}, and a comparison of 
the performance of these methods on this real world
data is detailed in Section~\ref{sec:realLISAComp}.

\subsection{Spatial Panel Models}

\label{sec:spModel}

The spatial panel data regression model 
\citep{ANSELIN2001,BALTAGI2003,KAPOOR2007,BALTAGI2008} extends panel 
data models into the spatial realm by accounting 
for both random region effects and
spatially autocorrelated residuals.  
This model can be fit to data via the 
\texttt{splm} R package \citep{SPLM}.

For $y_{t}$ being the observation vector from all regions 
at time $t$, we have
\begin{align}
\label{eqn:model1}
y_{t} &= \lmb W y_t + {X}_{t}\beta + u_{t}\\
\label{eqn:model2}
u_t &= \rho W u_t + \veps_t \\
\label{eqn:model3}
\veps_t &= \mu + v_t
\end{align}
from \cite{KAPOOR2007}. The model in \cite{BALTAGI2003}
is nearly identical but with a few swapped terms.
Equation~\ref{eqn:model1} models the observations based on 
the $n\times p$ matrix of non-stochastic predictors $X_{t}$
at time $t$
and unknown regression coefficients $\beta$. 
Additionally, $y$ can have a spatial lag determined by 
estimating $\abs{\lmb}<1$ and the user selected weight matrix $W$.
The $u_t$ in Equation~\ref{eqn:model2} is a spatially 
autocorrelated process depending on parameter $\abs{\rho}<1$. 
Equation~\ref{eqn:model3} models the innovations vector
as $\mu$, a regional random effect with variance $\sigma_\mu^2$, and 
$v_t$, iid mean zero normal 
errors with variance $\sigma_v^2$.
Within the \texttt{splm} package, the function 
\texttt{spgm} fits the above model using the generalized 
moments estimator from \cite{KAPOOR2007}
In contrast, the function \texttt{spml} 
fits a similar model using maximum 
likelihood as outlined in \cite{BALTAGI2003}.
Details on both methods can be found in 
\cite{SPLM}. In this article, we will fit the 
model defined by \ref{eqn:model1}, \ref{eqn:model2}, and
\ref{eqn:model3}
using \texttt{spgm}.

\subsection{Permutation Testing}

\label{sec:permTest}

There are two paradigmatic approaches to 
statistical hypothesis testing for spatial data models:
asymptotic normality and permutation testing.
The former method is popular due to the rapidity of 
producing a p-value, but relies on strong data assumptions
and a large enough sample size to be ``asymptotic''.
Much past research \citep{ANSELIN1995,ANSELIN2019,SEYAS2020}
suggests use of a permutation test instead of the 
normal approximation.  The permutation test
\citep{MIELKE2007,PESARIN2010,BROMBIN2013,GOOD2013}
is an exact non-parametric approach to statistic hypothesis
testing where the data is permuted in order to capture
the behaviour of a test statistic under the null 
hypothesis.  The biggest impediment to its universal
use is the computational burden of simulating massive 
numbers of permutations of one's dataset.
To solve this problem, the recent work of 
\cite{KASHLAK_KHINTCHINE2020} proposes an analytic
approach to computing p-values from permutation tests
for two-sample and $k$-sample testing for complex
data types like speech sounds.  Such analytic permutation
testing was extended to LISA and GISA statistics for univariate
spatial data in \cite{KASHLAK_YUAN_ABELECT}.
In this work, the previously investigated permutation testing 
framework is extended to spatial-temporal data specifically
for the spatial panel model.

\section{US County-wise Elections Data}
\label{sec:usData}

County-wise results for US presidential
elections are available online via the 
MIT Election Data and Science Lab \citep{USELECT_DATA}.
We will consider the electoral results over the 
five presidential elections from 2000 to 2016 inclusive.
For this analysis, the states of Alaska and Hawaii 
were removed so that the counties considered form a 
connected graph.  Also, three island counties were 
removed similarly as they have no edges in the graph;
these are Dukes and Nantucket county in Massachusetts
and San Juan county in Washington.  Lastly Broomfield county,
Colorado,
was removed from the dataset as it was incorporated
in 2001 and thus was not present for the election of 2000.

\begin{figure}
	\begin{center}
		\includegraphics[width=\textwidth]{\PICDIR/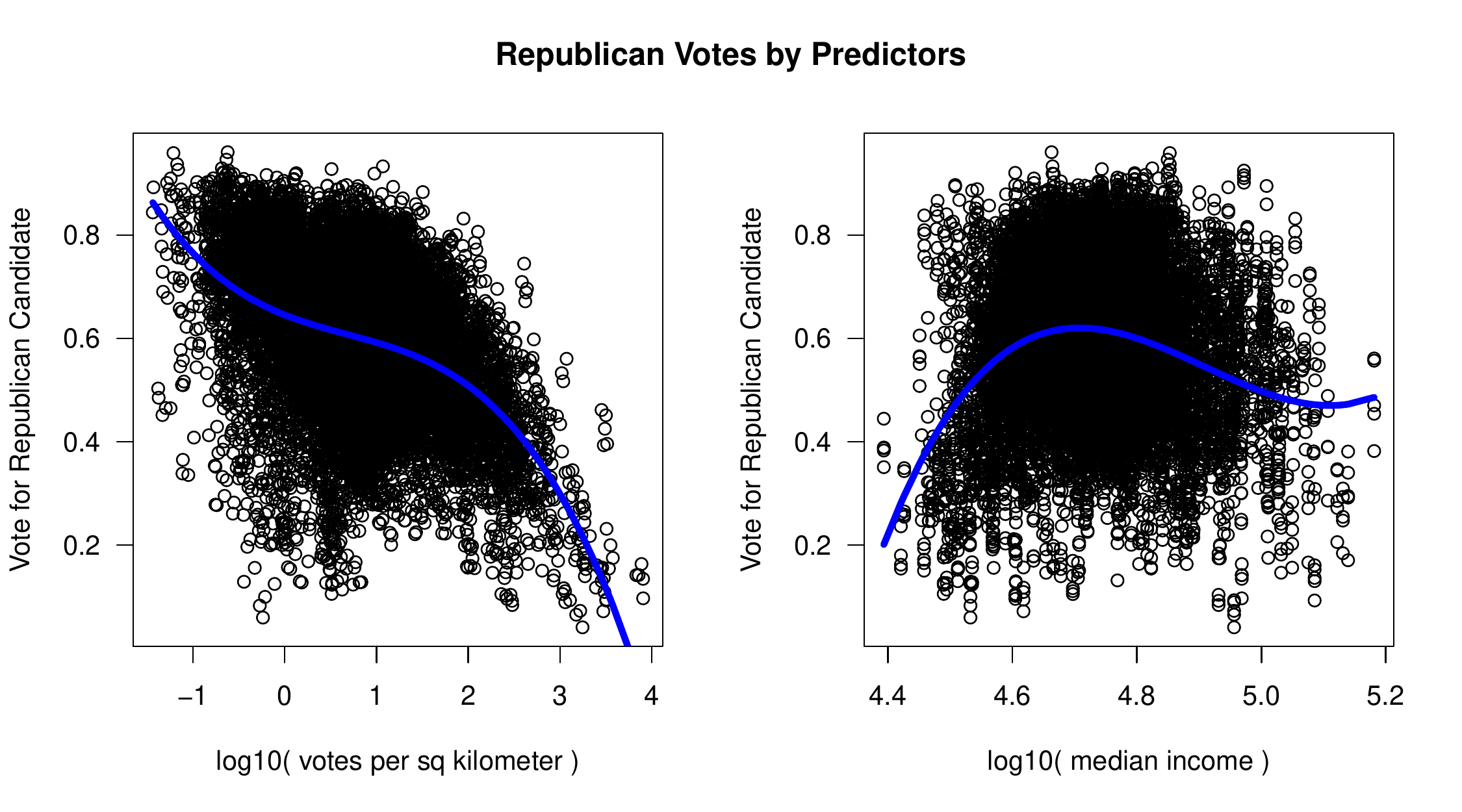}
		\includegraphics[width=0.5\textwidth]{\PICDIR/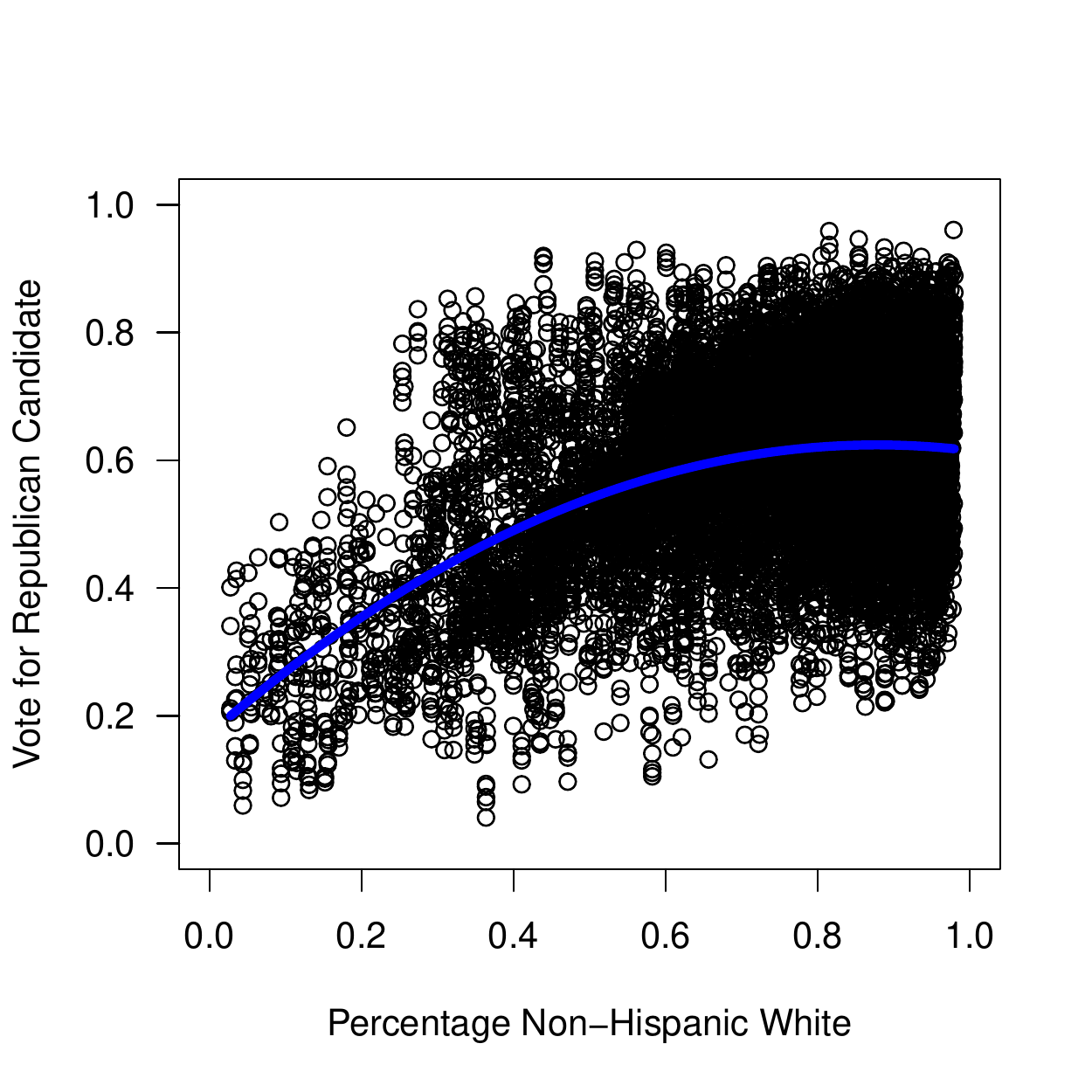}
	\end{center}
	\caption{
		\label{fig:repVote}
		The fraction of the vote for the Republican 
		candidate plotted against $\log_{10}$ voter
		population density (top left, $R^2=0.196$), $\log_{10}$ 
		median income (top right, $R^2=0.071$), 
		and percentage of the population that is
		non-Hispanic white (bottom, $R^2=0.143$).  
		The blue lines are 
		cubic polynomials fit by least squares for the
		top plots and a quadratic polynomial for the bottom.
	}
\end{figure}

The observations considered are the vectors $y_i\in[0,1]^5$ 
where $y_{ti}$ is the faction of the vote that went to
the Republican candidate; George W Bush; John McCain; 
Mitt Romney; and Donald Trump.
The predictor variables considered are 
the voter population density (log-scale), 
the state where the county resides (categorical),
median income (log-scale) downloaded 
from the Bureau of Labor Statistics, 
Local Area Unemployment Statistics,
and the percentage of the population that 
identifies as non-Hispanic white according to 
the U.S. Census Bureau's Population Division.
The relation between voting behaviour and each of the 
three continuous predictor variables is complex.
Hence, we consider linear, quadratic, and cubic 
polynomials for each of these.
Plots of these polynomials and the data are displayed
in Figure~\ref{fig:repVote}.  Comparing log voter density
to Republican vote, there is a noticeable negative trend 
indicating that the most densely populated counties
vote more heavily for the Democratic candidate. 
This coincides with the common assumption that dense cities
tend to vote left while sparse rural areas tend to vote right.
There is no strong positive or negative correlation
between median income and voter behaviour, but the 
cubic regression nevertheless identifies a drop in the
Republican vote for the poorest counties.
Lastly, a quadratic polynomial shows that the Republican
vote begins to drop on average as the non-Hispanic white
percentage of the population drops below 60\%.

\begin{figure}
	\begin{center}
		\includegraphics[width=\textwidth]{\PICDIR/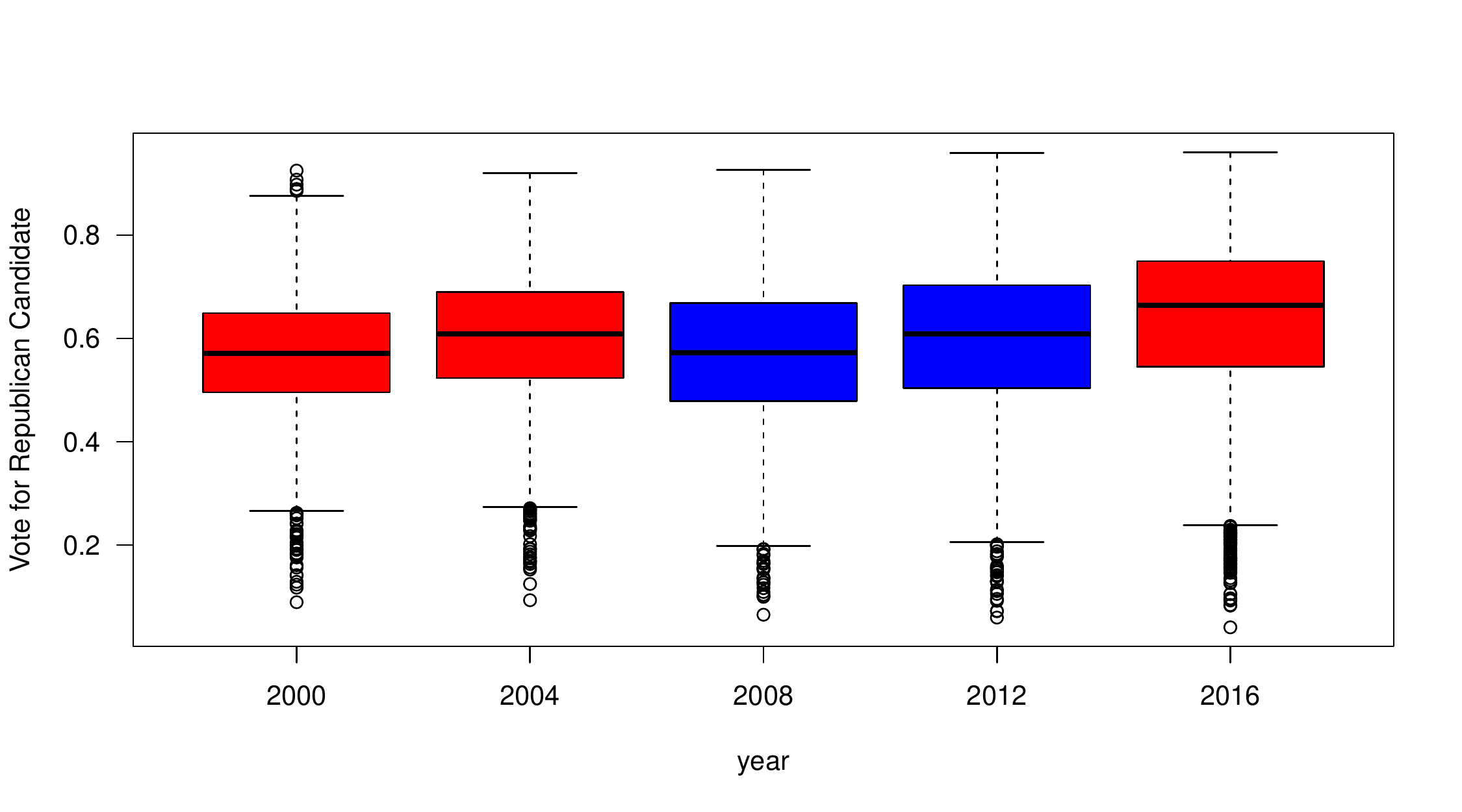}
	\end{center}
	\caption{
		\label{fig:repYear}
		Boxplots of the vote percentages for each year in the dataset
		coloured by the winning candidate (red = Republican, blue = Democrat).
	}
\end{figure}

\begin{figure}
	\begin{center}
		\includegraphics[width=\textwidth]{\PICDIR/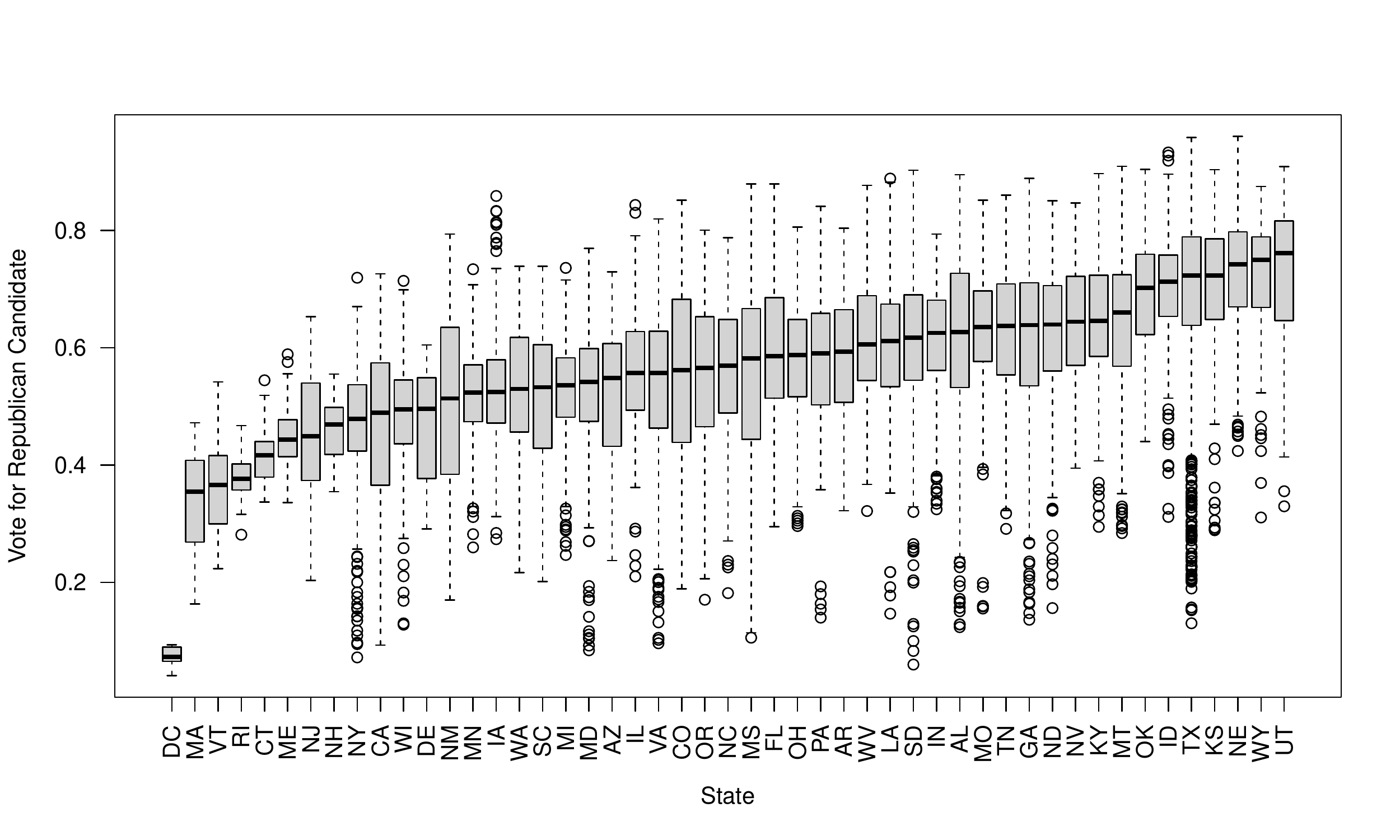}
	\end{center}
	\caption{
		\label{fig:repState}
		Boxplots of the vote percentages for each state in the dataset
		ordered from smallest to largest median Republican vote.
	}
\end{figure}

For categorical predictors, boxplots of the five election 
years are displayed in Figure~\ref{fig:repYear} and 
Table~\ref{tab:repYear} displays the results of a post-hoc 
Tukey test. The boxplots and table indicate that there is no
significant difference in the average countywise Republican
vote between years 2000 \& 2008 and 2004 \& 2012.  The 
average countywise vote was higher in 2016 for Donald Trump
than in the previous years.  However, Trump still received 
a smaller percentage of the popular vote than Hillary Clinton,
the Democratic challenger.
Due to the subtleties of the electoral college system used to 
elect a US president, a candidate's county-wise performance does 
not completely dictate the outcome of the election.  Aggregated over 
the five years of elections, the median county-wise Republican vote was 
60.1\% with 1st and 3rd quartiles of 50.5\% and 69.6\%.  However, 
the overall Republican popular vote---i.e. aggregated Republican votes over
all five elections divided by total votes cast---is only 47.4\%.
Boxplots are also plotted in Figure~\ref{fig:repState}
for each US state aggregated over all counties and the five elections.
These are ordered from smallest to largest median county 
Republican vote.

\begin{table}
	\caption{
		\label{tab:repYear}
		Results of a post-hoc Tukey test comparing election years
		with bolded entries having significant p-values with a test
		size of 5\%.
		The values in the table are column-year minus row-year.
	}
	\centering
	\begin{tabular}{lrrrr}
		\hline 
		& \multicolumn{4}{c}{Year}\\
		& 2004 & 2008 & 2012 & 2016 \\
		\hline
		2000 & {\bf3.29\%} & -0.10\% & {\bf2.67\%} & {\bf6.35\%} \\
		2004 & & {\bf-3.40\%} & -0.62\% & {\bf3.06\%} \\
		2008 & & & {\bf2.78\%}& {\bf6.46\%} \\
		2012 & & & & {\bf3.68\%}\\
		\hline
	\end{tabular}
\end{table}

\section{Local Spatial Association}

\subsection{LISA Statistics}

\label{sec:lisaStats}

Given spatial measurements or residuals from a fitted spatial
model, it is desirable to identify `hot spots' being any
spatial region where the observations are correlated.  
This led to the development of many Local Indicators of
Spatial Association (LISA) statistics
\citep{CLIFFORD1981,SOKAL1998,WALLER2004,GETISORDb,GAETAN2010,LUO2019,SEYAS2020}.
In what follows, we ignore normalizing constants as our 
entire approach centers around permutation testing which is
invariant to such constants.

For univariate data, we can define a few different LISA statistics.
Let $\mathcal{G}$ be a graph with $n$ vertices $\nu_1,\ldots,\nu_n$ 
and real valued measurements
$y_1,\ldots,y_n\in\real$ at each vertex.
For any choice of $n\times n$ weight matrix $W$,
Local Moran's index for vertex $i$ is
$$
I_i \sim \sum_{j=1}^n w_{i,j}(y_i-\bar{y})(y_j-\bar{y})
$$
where $w_{i,j}$ is the $i,j$th entry of $W$.
Local Geary's statistic for vertex $i$ is defined as
$$
C_i \sim \sum_{j=1}^n w_{i,j}(y_i-y_j)^2.
$$
The rank correlation statistic presented in  
\cite{KASHLAK_YUAN_ABELECT} and
referred to as binary association
is
$$
B_i \sim \beta_i \sum_{j=1}^n w_{i,j}\beta_j,~~
\beta_i = \left\{\begin{array}{rll}
1, & y_i \ge m \\
-1, & y_i < m
\end{array} \right.
$$
for $m = \mathrm{median}(y)$.  
The classic Gettis-Ord statistics are not included here as they 
are equivalent to Moran's index under the permutation 
methodology.

For multivariate data, 
let $y_i\in\real^T$ be a $T$-long vector for the $i$th vertex.
These univariate LISA statistics can be extended to this setting 
as follows.
For a fixed symmetric positive definite $T\times T$ matrix $M$,
an inner product on $\real^T$ can be defined as 
$\iprod{a}{b}_M = \TT{a}Mb$ for any $a,b\in\real^T$.
Thus, for any such choice of $M$, the vector version of
Moran's statistic is 
$$
I_i \sim \sum_{j=1}^n w_{i,j} \iprod{y_i-\bar{y}}{y_j-\bar{y}}_M
$$
where we again ignore scaling constants.
For $Y = ( y_1~y_2~\ldots~y_n )\in\real^{T\times n}$, this statistic
can be quickly computed at every vertex via the formula
$$ 
I = \mathrm{diag}\left[\TT{(Y-\bar{Y})}M(Y-\bar{Y}) W\right] \in \real^n
$$
where $\mathrm{diag}[A]$ extracts the diagonal of matrix $A$ and
$\bar{Y} = ( \bar{y}~\bar{y}~\ldots~\bar{y} )$.
Geary's statistic can be extended to multivariate data by
simply considering 
$$
C_i \sim \sum_{i=1}^n w_{i,j}\norm{ y_i - y_j }_{\ell^p}^p
$$
for some choice of $\ell^p$ norm.  For the Euclidean norm,
this can be computed for all vertices quickly as 
$
\norm{y_i - y_j}_{\ell^2}^2 =
\norm{y_i}_{\ell^2}^2 + \norm{y_j}_{\ell^2}^2
- 2\iprod{y_i}{y_j}.
$
Noting that the matrix $\TT{Y}Y$ has $i,j$th entry $\iprod{y_i}{y_j}$
and setting $d = \mathrm{diag}(\TT{Y}Y)$,
$$
C = \mathrm{diag}\left\{[ (d\oplus d) - 2\TT{Y}Y]W\right\} \in \real^n
$$
where $(d\oplus d)_{i,j} = d_i + d_j$.
Lastly, the binary association statistic can be simply 
generalized to the multivariate setting in the same manor 
as with Moran's index:
$$
B_i \sim \sum_{j=1}^n w_{i,j}\iprod{\beta_i }{\beta_j},~~
\beta_{i,k} = \left\{\begin{array}{rll}
1, & y_{i,k} \ge m_k \\
-1, & y_{i,k} < m_k
\end{array} \right.
$$
for $m_k = \mathrm{median}( y_{i,k}\,:\, i=1,\ldots,n )$.
For binary association, we just consider the standard 
Euclidean inner product, i.e. the dot product.

\subsection{Significance Testing}
\label{sec:sigTest}

\subsubsection{Local Testing}

A general approach to testing LISA statistics by 
analytically bounding the permutation test p-value
is introduced in \cite{KASHLAK_YUAN_ABELECT}.  Here, 
we extend this work to the multivariate setting.

The gamma index \citep{MANTEL1967,HUBERT1985}
is a general measure of matrix similarity,
$
\gamma_{AB} := \sum_{i,j=1}^na_{i,j}b_{i,j}
$,
for two similar matrices $A$ and $B$.
For specific choices of $A$ and $B$, the gamma index
can be treated as a general correlation statistic.
The local gamma index \citep{ANSELIN1995} is similarly
defined as 
$
\gamma_i = \sum_{j=1}^n a_{i,j}b_{i,j}.
$
The formulae from Section~\ref{sec:lisaStats} can be rewritten
in terms of a local gamma index by choosing $A$ to be the weight
matrix $W$ and $B$ to be one of the entries in Table~\ref{tab:locGam}.
To align with the notation of \cite{ANSELIN1995} and 
\cite{KASHLAK_YUAN_ABELECT}, we will write $a_{i,j} = w_{i,j}$,
$b_{i,j} = \lmb( y_i, y_j )$ for some similarity function
$\lmb$, and finally that
$\gamma_{i} = \sum_{j=1}^n w_{i,j}\lmb( y_i, y_j )$.
Using this notation, we can define the permutated 
local gamma index to be 
$
\gamma_{i}(\pi) = \sum_{j=1}^n w_{i,j}\lmb( y_i, y_{\pi(j)} )
$
where $\pi$ is a uniformly random element of $\mathbb{S}_{n}$, the 
symmetric group on $n$ elements, conditioned so that $\pi(i)=i$.

\begin{table}
	\caption{
		\label{tab:locGam} 
		Choices of matrices $A$ and $B$ to get a $\gamma_i$ equivalent
		to one of the LISA statistics.
	}
	\centering
	\begin{tabular}{lcc}
		\hline
		LISA & $A_{i,j}$ & $B_{i,j}$ \\ \hline
		Moran & $w_{i,j}$& $\iprod{y_i-\bar{y}}{y_j-\bar{y}}_M$\\
		Geary & $w_{i,j}$& $\norm{ y_i - y_j }_{\ell^p}^p$\\
		Binary& $w_{i,j}$& $\iprod{\beta_i }{\beta_j}$\\
		\hline
	\end{tabular}
\end{table}

In \cite{KASHLAK_YUAN_ABELECT}, it is proven that for any such gamma
index constructed with a binary weight matrix---i.e. $w_{i,j} = 0,1$
for all $i,j=1,\ldots,n$---the following concentration inequality holds
when $m_i = \sum_{j=1}^nw_{i,j} \ll n/2$ or $m_i \gg n/2$,
\begin{equation}
\label{eqn:tailLocal}
\prob{ \abs{\gamma_i{(\pi)}-m_i\bar{\lmb}_{-i}} \ge \gamma_i \mid y_1,\ldots,y_n } \le 
\frac{1}{\sqrt{\pi}}\Gamma\left(
\frac{n-1}{m_i(n-m_i-1)}\frac{\gamma_i^2}{2s_i^2}\verb|;|\frac{1}{2}
\right) + O(n^{-4})
\end{equation}
where $\mathrm{P}(~)$ is the uniform probability measure on the 
symmetric group conditioned so that $\pi(i)=i$, 
$\bar{\lmb}_{-i} = (n-1)^{-1}\sum_{j=1}^n\lmb(y_i,y_j)\indc{i\ne j}$,
$\Gamma(~)$ is the upper incomplete gamma function, and 
$s_i^2 = (n-1)^{-1}\sum_{j\ne i}(\lmb_{i,j}-\bar{\lmb}_{-i})^2$
is the sample variance of the $i$th row.

\begin{remark}
	Typically, for planar data such as that considered in this 
	work, $m_i\ll n/2$ for all vertices $i=1,\ldots,n$.  However,
	in the case where $m_i\sim n/2$, 
	we define 
	\begin{align*}
	\varpi_- &= {\min\{m_i,n-m_i-1\}}/{\max\{m_i,n-m_i-1\}^2}, \text{ and}\\
	\varpi_+ &= {\max\{m_i,n-m_i-1\}}/{\min\{m_i,n-m_i-1\}^2}.
	\end{align*}
	Then, the
	concentration inequality becomes
	$$
	\prob{ \abs{\gamma_i{(\pi)}-m_i\bar{\lmb}_{-i}} \ge \gamma_i \mid
		y_1,\ldots,y_n } 
	\le  C_0 I\left[
	\exp\left( 
	-\frac{\gamma_i^2}{2s_i^2}\varpi_-
	\right)\verb|;|
	(n-1)\varpi_+
	,\frac{1}{2}
	\right]
	$$
	where $I[\cdot]$ is the regularized incomplete beta function,
	and 
	$
	C_0 = \frac{
		{\sqrt{(n-1)\varpi_+}}\Gamma\left((n-1)\varpi_+\right)
	}{
		\Gamma\left(\frac{1}{2}+(n-1)\varpi_+\right)
	}
	$
	with $\Gamma(\cdot)$ the (complete) gamma function. 
\end{remark}

\subsubsection{Global Testing}

Statistics for local indicators of spatial association can be 
extended to statistics for global indicators of spatial association
(GISA).  In \cite{KASHLAK_YUAN_ABELECT}, a concentration inequality
similar to that for LISA statistics is proved.  Given the 
same setup as the previous section, then
\begin{equation}
\label{eqn:tailGlobal}
\prob{ 
	\abs*{\gamma(\boldsymbol{\pi})- \sum_{i=1}^n m_i\bar{\lmb}_{-i} } \ge \gamma 
	\,\mid\, y_1,\ldots,y_n
}
\le
\frac{1}{\sqrt{\pi}}\Gamma\left(
\frac{\gamma^2}{4\upsilon^2}
\verb|;|\frac{1}{2}
\right) + O(2^{-2n})
\end{equation}
where $\boldsymbol{\pi}=(\pi_1,\ldots,\pi_n)$
with $\pi_i(i)=i$,
$
\gamma(\boldsymbol{\pi}) = \sum_{i=1}^n \gamma_i(\pi_i)
$
is the permuted variant of this test statistic,
and
$\upsilon^2=\sum_{i=1}^n \eta_i s_i^2$ with
$\eta_i = m_i(n-m_i-1)/(n-1)$.

The GISA tests apply to the same statistics as
the LISA tests do. These GISA test statistics are
more directly comparable  with the LM and LR tests 
considered in \cite{BREUSCH1980,ANSELINBERA1998,BALTAGI2003,HSIAO2014}
and others.

\section{Simulation Studies}

\label{sec:simStudy}

\subsection{LISA Tests}

Four LISA statistics---Moran, Geary $\ell^2$ and $\ell^1$,
and Binary association---are tested on simulated data 
on a $(50\times60)$-vertex rectangular grid where up to 
four edges exist for each vertex connecting it to those 
vertices above and below and to the left and to the right.
Denoting the graph adjacency matrix as $A$, we 
generate $200$ replicates of mean zero 
$5\times 3000$ dimensional 
multivariate Gaussian data
with covariance $I + cA$ where $c\in[-0.25,0.25]$.

\begin{figure}
	\begin{center}
		\includegraphics[width=0.475\textwidth]{\PICDIR/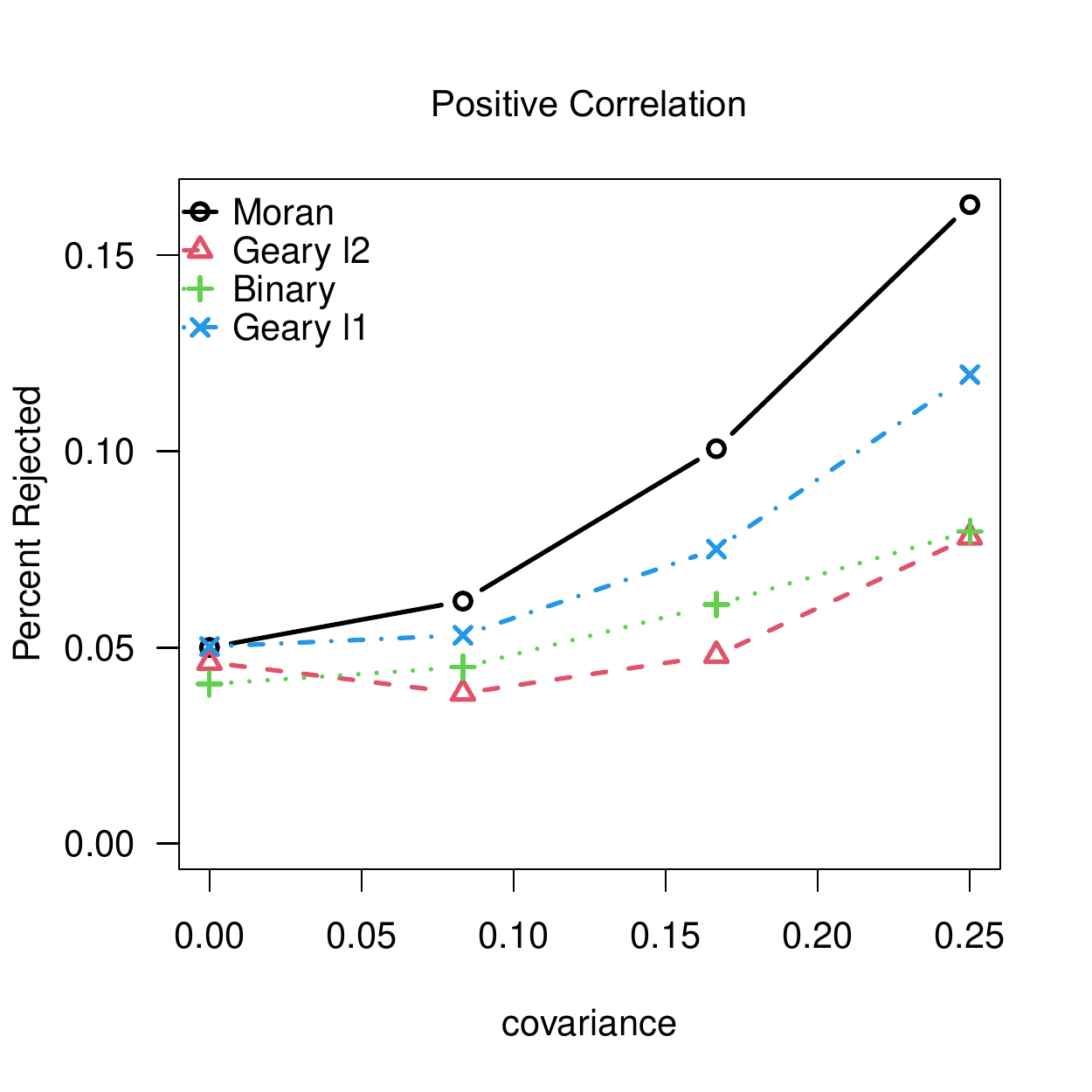}
		\includegraphics[width=0.475\textwidth]{\PICDIR/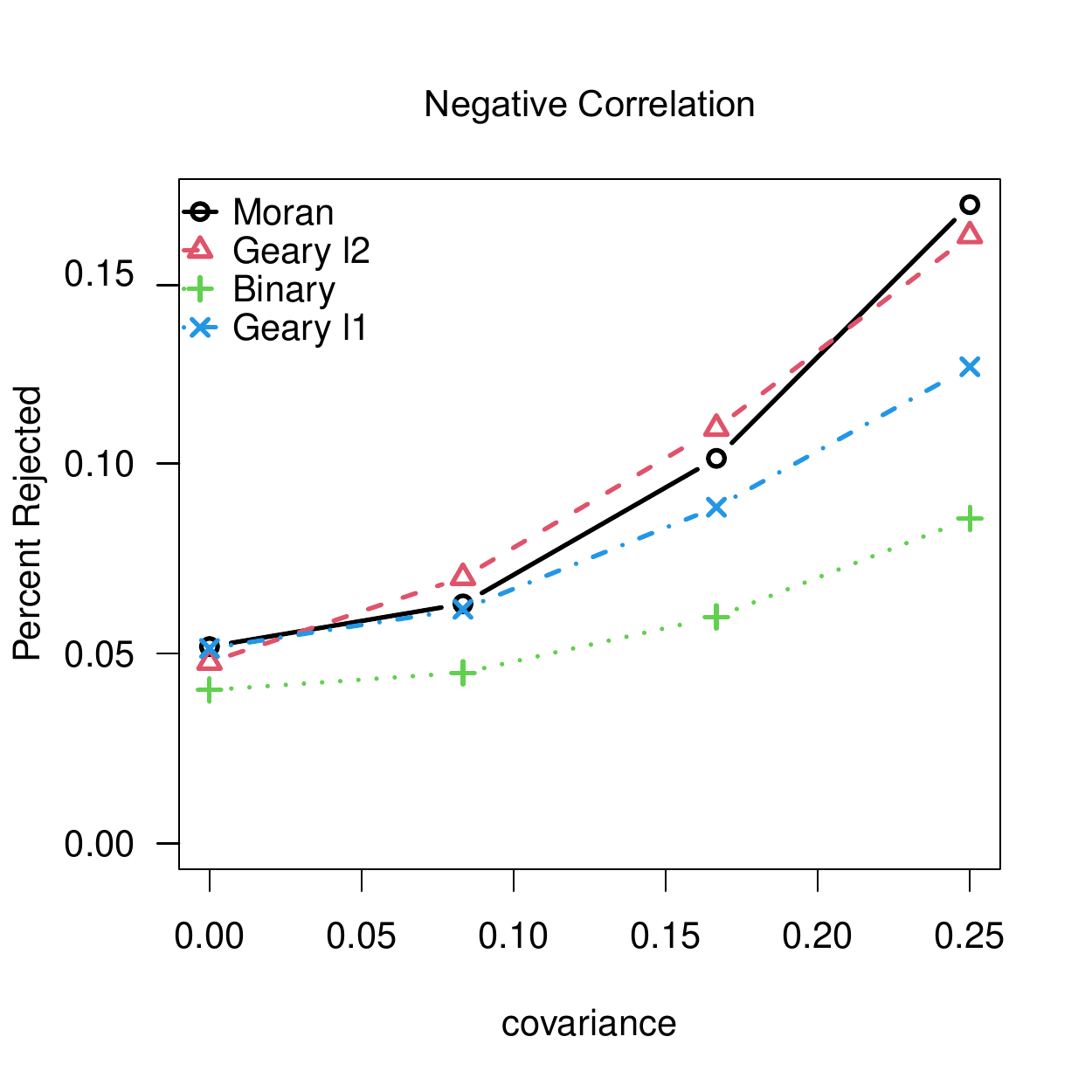}
	\end{center}
	\caption{
		\label{fig:sim1}
		Power curves for the four LISA tests as the covariance
		increases in the positive direction
		(left) and negative direction (right).
	}
\end{figure}

Figure~\ref{fig:sim1} charts the number of vertices 
with p-values less than 5\% for each of the four LISA
statistics with p-values computed via formula~\ref{eqn:tailLocal}.
As $c$ increases from zero, Moran's statistic is shown
to identify the most significant vertices followed by 
Geary with the $\ell^1$ norm. 
Geary $\ell^2$ and binary association perform 
the worst.
In contrast, for $c$ decreasing invoking negative
correlations between adjacent vertices, the power curves
for Moran and Geary $\ell^2$ show similar performance
with Geary $\ell^1$ identifying fewer significant
vertices and with binary association performing the worst.

In these simulations, the binary association statistic
performed the worst in both testing scenarios.  
However, in the next section, it is shown to have 
good performance on the US elections data while  Geary with
the $\ell^2$ norm identified the fewest hot spots.  Moran's
statistic consistently achieves the highest power
in all testing scenarios.  However, the other three
methods often identify vertices missed by Moran's 
statistic.  This suggests that these LISA statistics
can be used to complement one another in an exploratory analysis.

\subsection{GISA Tests}

\begin{figure}
	\begin{center}
		\includegraphics[width=0.475\textwidth]{\PICDIR/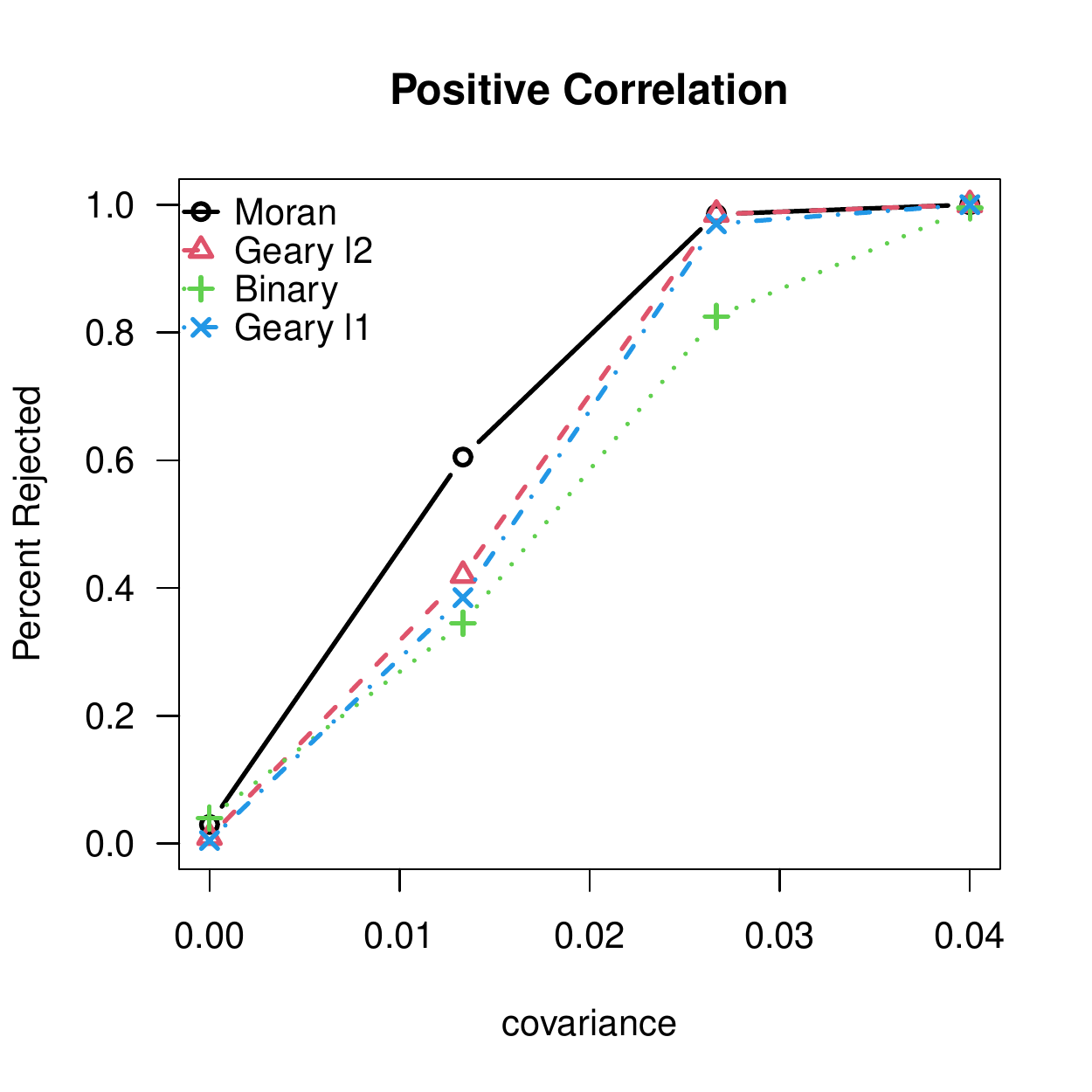}
		\includegraphics[width=0.475\textwidth]{\PICDIR/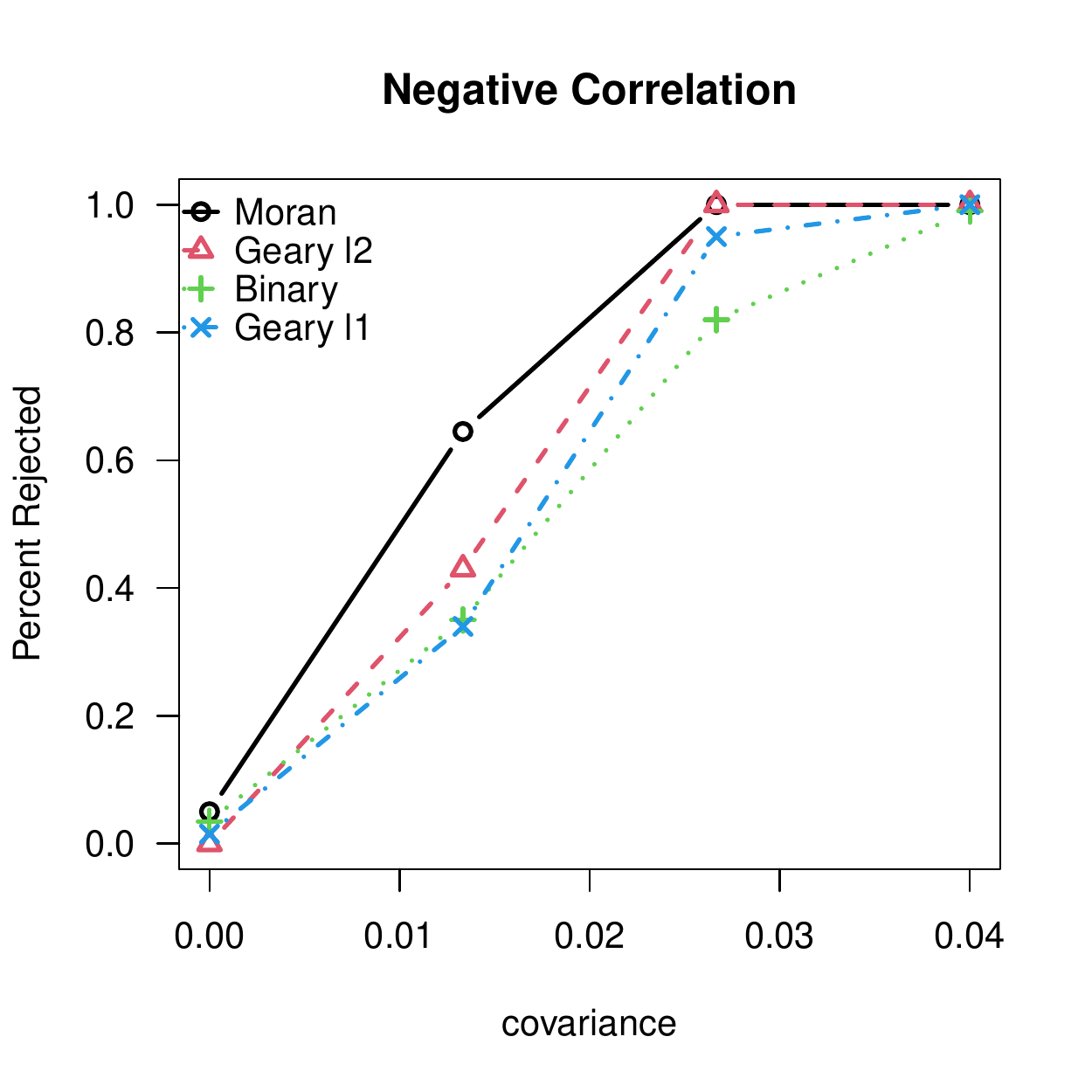}
	\end{center}
	\caption{
		\label{fig:sim2}
		Power curves for the four GISA tests as the covariance
		increases in the positive direction
		(left) and negative direction (right).
	}
\end{figure}

Testing for global spatial association (GISA) is 
also possible using formula~\ref{eqn:tailGlobal}.
Computing p-values for the same simulated data and
test statistics considered above results in 
Figure~\ref{fig:sim2}.
For positively correlated neighbours, the global 
version of Moran's test gives the best statistical
power.  Geary's $\ell^2$ and then $\ell^1$ statistics
are next, and the global binary association test
performs the worst, but still is able to achieve
a good amount of power to identify the presence of 
global spatial association.
Similar behaviour is seen for a negatively correlated
network.  Of note, Geary's $\ell^2$ global test gives 
slightly higher power than Geary's $\ell^1$ whereas
for local testing, Geary with the $\ell^1$ norm
has stronger performance than $\ell^2$ for 
positively correlated data.

In the real data sections to follow, we do not 
discuss GISA testing as all of the p-values are 
extremely small for all four statistics and all 
five neighbourhoods considered.  This is evident in the LISA
plots displayed below in Figure~\ref{fig:lisaMaps},
which show many areas of spatial association 
under each of the four statistics considered.

\section{Real Data Results}

\subsection{Spatial Panel Data Regression}
\label{sec:realSPLM}

Fitting and testing a spatial panel model requires 
selection of a weight matrix. To coincide with the 
theorems in \cite{KASHLAK_YUAN_ABELECT}, we will 
only consider binary weight matrices---i.e. those
with entries 0 and 1 only.  Among such weight matrices,
we denote $W_k$ to be the $k$-lagged weight matrix.
Beginning with  $W_0=I$ and $W_1$ being the graph
adjacency matrix, we recursively define
$$
W_k = W_1^k\prod_{l=0}^{k-1}(\boldsymbol{1}-W_l)
$$
where $\boldsymbol{1}$ is the $n\times n$ matrix of
all ones.
We will consider neighbourhoods for $k=1,\ldots,5$.

The fitted spatial panel models are compared in 
Table~\ref{tab:spmCoefs}.  Both $\lmb$,
the spatial lag parameter,  and $\rho$,
the spatial autocorrelated errors parameter, 
decrease in magnitude as the lag increases.
Thus, the spatial dependence in the data 
begins to wane as the neighbourhood spreads out.
Thus, in the next section, 
we will focus on testing for local spatial association
with the adjacency weight matrix.  The 
$R^2$ values for the five fitted models all are 
around 60\%.

Fitting a spatial panel data regression model to this 
data captures the behaviour of voting in most US counties.
However, extreme counties with respect to the predictors 
result in inaccurate or erroneous predictions.  
On the political left, the counties of New York City and the 
Bronx both have fitted values less than zero.  NYC 
has the highest voter density with between 7000 and 8000 
voters per square kilometre across the five elections.
The next densest county is King's County, NY with 
approximately 3500 voters per square kilometer.
The Bronx is also one of the counties with the 
densest population, but while NYC is about 47\% white
and King's is 37.5\% white, the Bronx is less than
10\% white.
On the political right, both King and Loving county, Texas,
get fitted values greater than 100\%.  This is mainly 
due to being very sparsely populated 
($\sim$ 1 voter per 20 square kilometers )
and in Texas, which gives a boost to the expected
Republican vote as discussed next.

\begin{table}
	\caption{
		\label{tab:spmCoefs}
		A comparison of spatial panel models for 
		neighbourhoods defined by lags 1,\ldots,5.
	}
	\centering
	\begin{tabular}{lrrrrr}
		\hline
		& \multicolumn{5}{c}{Lag} \\
		& 1 & 2 & 3 & 4 & 5 \\
		\hline
		$\lmb$ & 0.0225 & 0.0094 & 0.0035 & 0.0018 & 0.0013 \\
		$\rho$ & -0.0779& -0.0180& 0.0051 & 0.0086 & 0.0088 \\
		$\sigma_v^2$   & 0.0024 & 0.0023 & 0.0024 & 0.0023 & 0.0023\\
		$\sigma_\mu^2$ & 0.0242 & 0.0238 & 0.0238 & 0.0238 & 0.0239\\
		$R^2$ & 59.4\% & 62.2\% & 62.9\% & 63.1\% & 63.1\% \\
		\hline
	\end{tabular}
\end{table}

\begin{figure}
	\begin{center}
		\includegraphics[width=0.31\textwidth]{\PICDIR/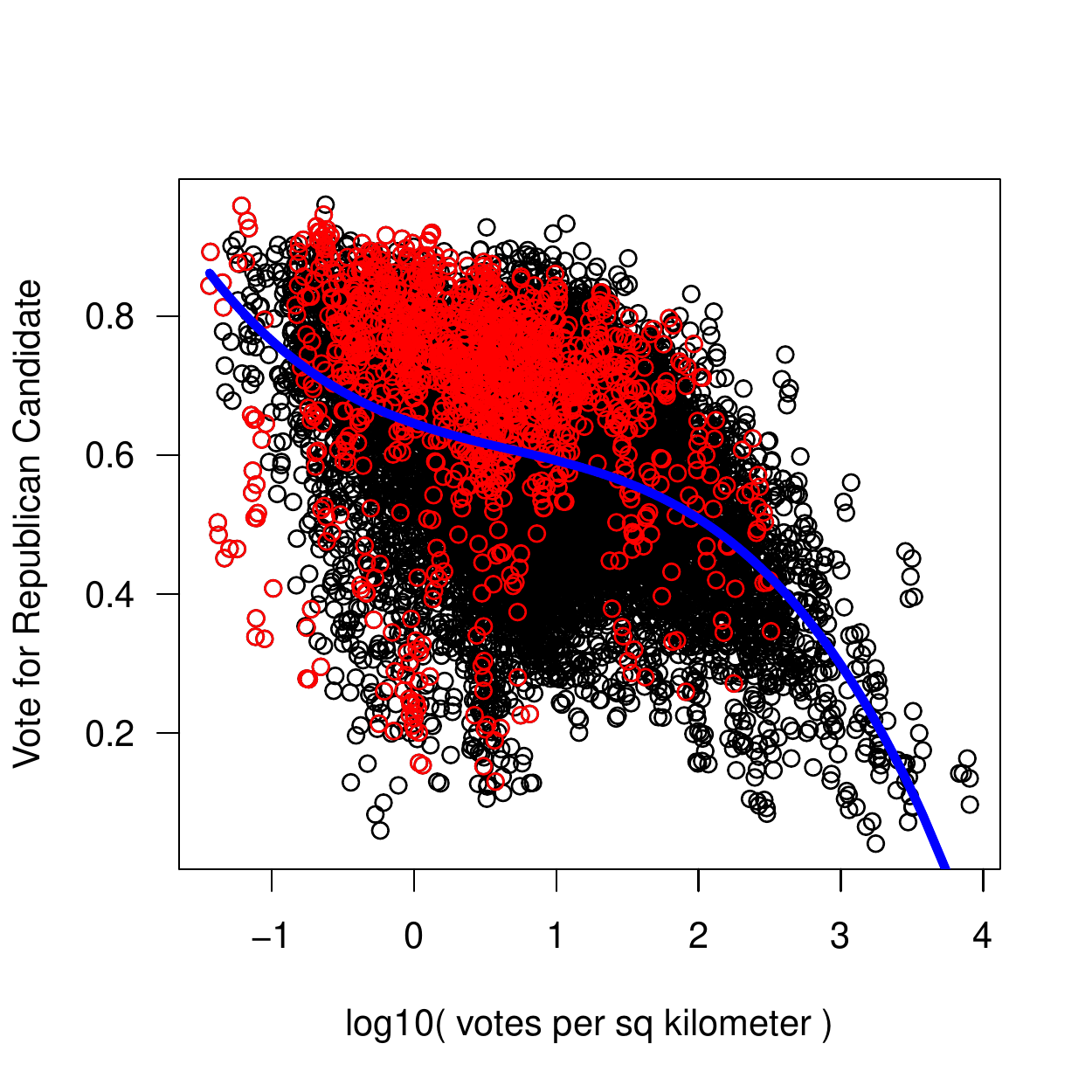}
		\includegraphics[width=0.31\textwidth]{\PICDIR/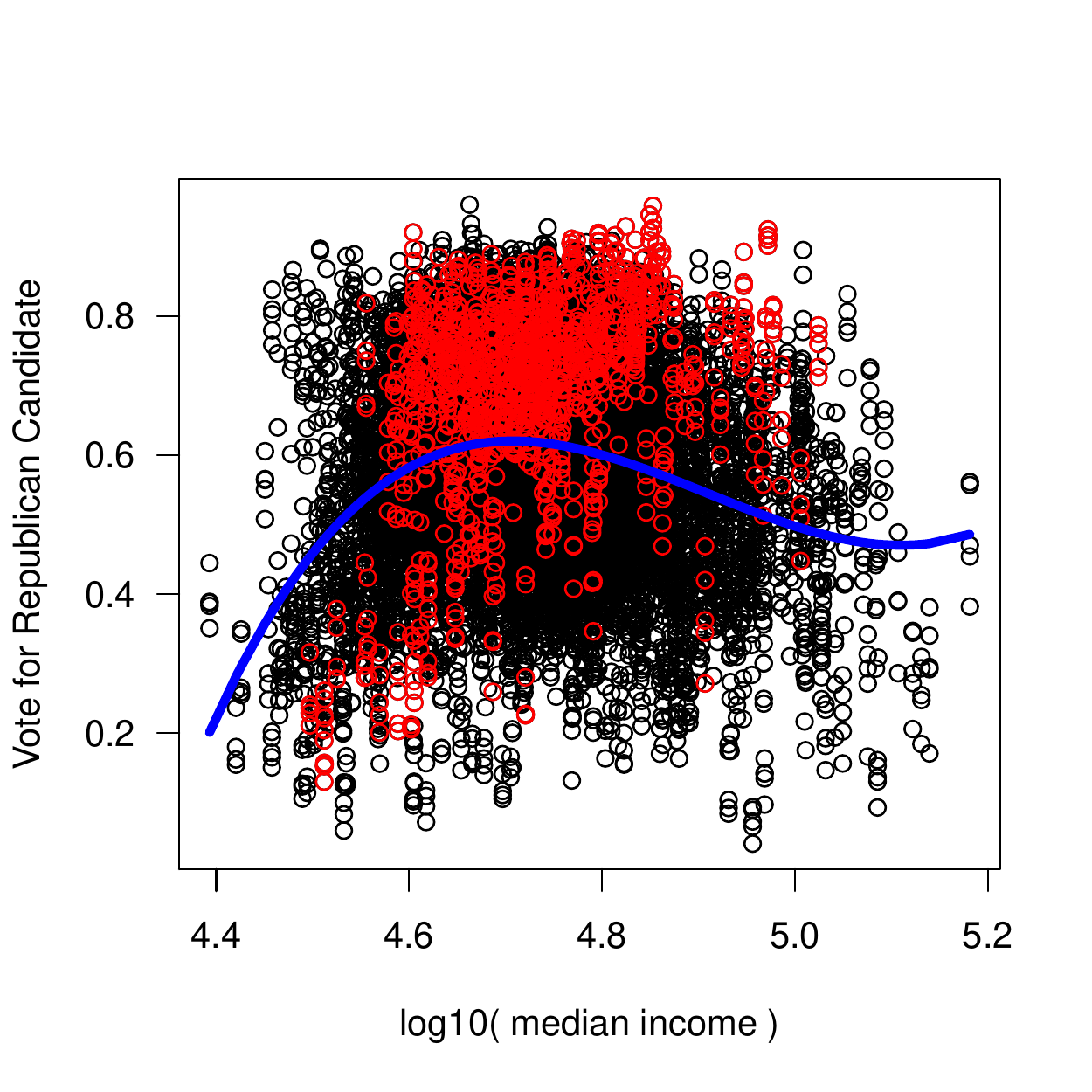}
		\includegraphics[width=0.31\textwidth]{\PICDIR/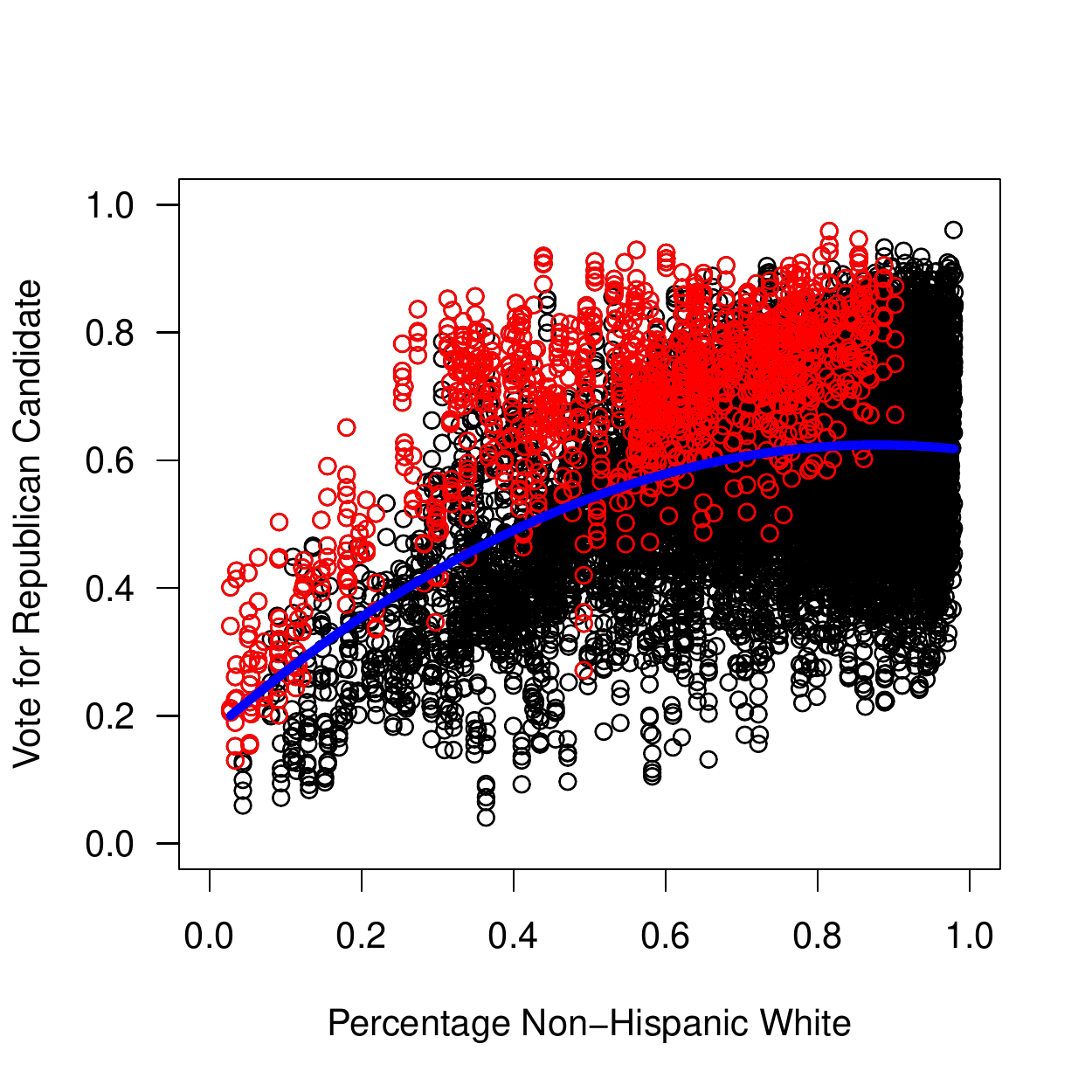}
	\end{center}
	\caption{
		\label{fig:repVoteTx}
		A copy of Figure~\ref{fig:repVote} with the Texas
		counties coloured in red. This demonstrates a higher
		percentage of Texan vote going towards the Republican
		candidate than expected by the three predictors.
	}
\end{figure}

Inclusion of the state as a categorical predictor 
allows us to identify those states that vote more or less
favourably for the Republican candidate than the three continuous 
predictors---population density, median income, and non-Hispanic 
whiteness---would 
suggest.  Most notably, the state of Texas has an estimated
upward shift of 19.2\% indicating that Texan counties on 
average vote more heavily for the Republican candidate than 
expected given the other predictors.  This is visualized
in Figure~\ref{fig:repVoteTx}, which shows the fitted regression
lines for the three predictors with Texan counties coloured in 
red.  The majority of Texan counties lie above the regression 
lines.

\subsection{Local Spatial Association}

\label{sec:realLISA}

Four multivariate LISA statistics are applied to the 
residuals for the fitted models.  These are Moran's
$I$, Geary's statistic in both $\ell^2$ and $\ell^1$,
and binary association.
The p-values produced by formula~\ref{eqn:tailLocal}
are subsequently adjusted for multiple testing using
the Benjamini-Hochberg method for False Discovery Rate 
(FDR) control 
within the spatial framework making use of 
the \texttt{p.adjustSP} function in the 
\texttt{spdep} R package 
\citep{SPDEP1,SPDEP2}.
The p-values reported were computed with $W_1$, 
the adjacency matrix, as the chosen weight matrix.
Similar results were seen for other weight matrices.

\begin{figure}
	\begin{center}
		\includegraphics[width=0.45\textwidth]{\PICDIR/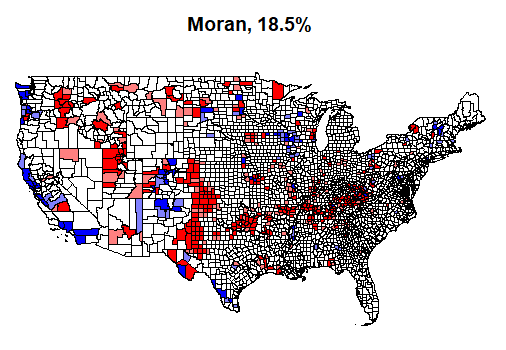}
		\includegraphics[width=0.45\textwidth]{\PICDIR/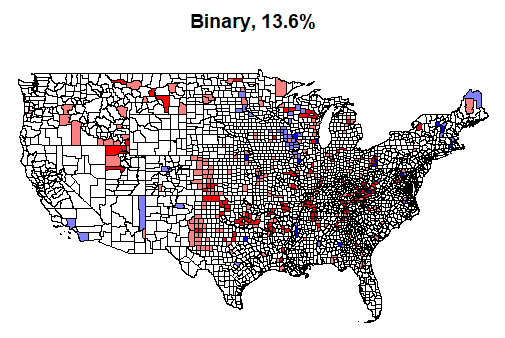}
		\includegraphics[width=0.45\textwidth]{\PICDIR/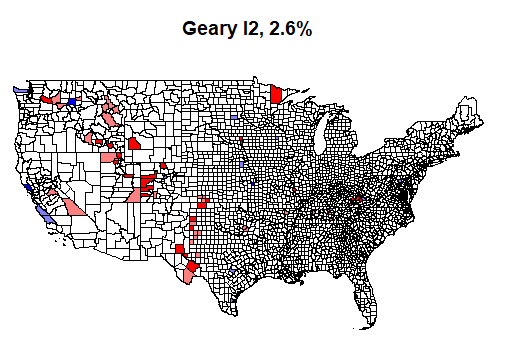}
		\includegraphics[width=0.45\textwidth]{\PICDIR/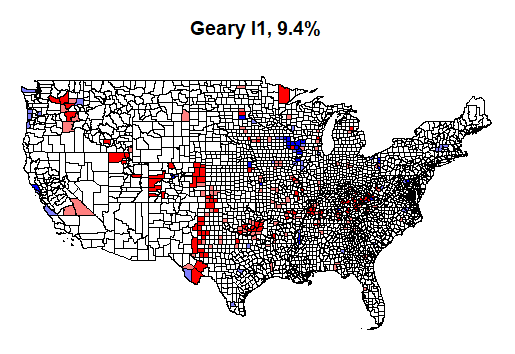}
	\end{center}
	\caption{
		\label{fig:lisaMaps}
		Maps displaying significant counties with respect to a given 
		LISA statistic after adjusting for multiple testing.  
		The percentage of significant counties in displayed
		in the plot titles.
	}
\end{figure}

Counties with significant p-values under each of the 
LISA statistics are displayed in Figure~\ref{fig:lisaMaps},
which are coloured by red or blue depending on whether the
Republican or Democratic candidate took more than 50\% of 
the county's vote in three or more of the years considered.
This allows for the visualization of hot spots where the 
residuals all trend in a similar direction.
Moran's statistic finds the most significant counties
of the four methods; 18.5\% of the 3104 counties are designated
as significant at an FDR of 5\%. Binary association 
comes next at 13.6\%, Geary's statistic with the $\ell^1$
norm at 9.4\% and lastly Geary's statistic with $\ell^2$
with only 2.6\% of the counties deemed to have significant
spatial association after correcting for multiple testing.
The lighter colours indicate significant  at 5\% FDR and
the darker colours indicate significant at 1\% FDR.

LISA statistics can detect both significant positive and 
negative spatial association.  However, nearly all of the
counties detected in Figure~\ref{fig:lisaMaps} are due to
positive association.
As these statistics are computed on the residuals 
produced from the spatial panel model, positive
spatial association implies the existence of a 
geographic region where the residuals trend in the 
same direction.  These are geographic voting blocks
whose votes trend in a similar direction even after
the three predictors and US state are taken into account.
One of the most noticeable correlated collections of
counties is the Republican voting 
north-south region spanning from west Texas
upward through the Kansas-Colorado border.
Another Republican voting block exists around 
the boarder of Kentucky with Virginia and West Virginia
extending into Tennessee.
The Moran, binary, and $\ell^1$ Geary maps highlight a 
Democrat voting region along the Mississippi river
as it separates the states of Iowa and Minnesota 
from Illinois and Wisconsin.  The Moran map also 
detects significant spatial autocorrelation 
along the left-voting west coast counties.

\subsection{Comparison of LISA Statistics}

\label{sec:realLISAComp}

In the previous section, Figure~\ref{fig:lisaMaps}
shows that the number of significant counties detected
can vary a lot between the four methods.  
To further contrast the four methods, we can consider 
$2\times2$ tables for the agreements and disagreements
between each pair of methods.
In Table~\ref{tab:tableComp}, the Matthews Correlation
Coefficient, 
$$
\text{MCC} = \frac{
	TP \times TN - FP \times FN
}{
	\sqrt{
		(TP+FP)(TP+FN)(TN+FP)(TN+FN)
	}
},
$$
and the Rand index,
$$
\text{Rand} = \frac{
	TP + TN
}{
	TP+FP+FN+TN
},
$$
are computed for each pairing of methods. 
For simplicity of notation, we use true/false
positive/negative to refer to the methods 
agreeing or disagreeing on which counties 
are chosen have significant spatial association.

The Rand index or percentage of agreement is 
above 80\% for each pairing of methods.  This
is due to most of the counties being deemed 
non-significant (TN) by all methods.  The
MCC gives a more nuanced comparison of the 
methods.  Higher MCCs are seen for Moran and
Binary association, both of which are of the 
form of an inner product, and for 
Geary $\ell^1$ and $\ell^2$, both of which
are of the form of an $\ell^p$ norm.

\begin{table}
	\caption{
		\label{tab:tableComp}
		A comparison of the four LISA statistics using
		Matthews  Correlation (left) and using 
		the Rand Index (right).
	}
	\centering
	\begin{tabular}{lccccccc}
		\hline
		&\multicolumn{3}{c}{\bf MCC}&&\multicolumn{3}{c}{\bf Rand}\\
		& Geary $\ell^2$ & Geary $\ell^1$ & Binary & 
		& Geary $\ell^2$ & Geary $\ell^1$ & Binary\\
		\hline
		Moran &          0.232&0.437&0.447&&0.828&0.855&0.849\\
		Geary $\ell^2$&       &0.394&0.132&&     &0.860&0.922\\
		Geary $\ell^1$&       &     &0.412&&     &     &0.878\\
		\hline
	\end{tabular}
\end{table}

\section{Discussion and Extensions}

There are many ways to test for local and global
spatial association for spatial panel data models.
In this work, we have seen that the extension of 
Moran's statistic to the multivariate domain gives
the best power to identify local clusters of spatially
dependent regions.  However, other methods have both 
comparable statistical power and typically identify
different significant regions.  Thus, an ensemble 
approach to detecting spatially dependent regions 
is warranted.

The extensions of LISA statistics presented in 
Section~\ref{sec:lisaStats} are based on inner 
products (Moran and binary association) and on 
$\ell^p$ norms (Geary).  This naturally can be
further applied to spatially observed functional 
data such as climate data, linguistic data, and others
\citep{DELICADO2010,MENAFOGLIO2016,TAVAKOLI2019}.
Panel data, longitudinal data, functional data, 
and time series data are all closely related objects
of study that often are collected from a 
spatial domain.  Our methodology  is extendable to 
such areas of analysis.

\section*{Acknowledgements}

The authors would like to thank the Natural Sciences 
and Engineering Research Council of Canada (NSERC) for
their funding support.





\bibliographystyle{plainnat}
\def\BIBDIRWIN{.}
\bibliography{\BIBDIRWIN/kasharticle,\BIBDIRWIN/kashbook,\BIBDIRWIN/kashpack,\BIBDIRWIN/kashself}



\end{document}